\documentclass[a4paper,11pt]{article}

\usepackage[utf8]{inputenc}
\usepackage[T1]{fontenc}
\usepackage{lmodern}

\usepackage{amsmath, amssymb}
\usepackage{siunitx}
\usepackage{graphicx}
\usepackage{subcaption}
\usepackage{float}

\usepackage{comment}
\usepackage{geometry}
\usepackage{booktabs}

\usepackage{hyperref}
\usepackage{cite}

\title{Scalable Relay Switching Platform for Automated Multi-Point Resistance Measurements}

\author{
Edoardo Boretti,
Kostiantyn Torokhtii,
Enrico Silva,
Andrea Alimenti*\\[0.5em]
\small Department of Industrial, Electronic and Mechanical Engineering\\
\small Roma Tre University, Italy\\
\small \texttt{*andrea.alimenti@uniroma3.it}
}

\date{}

\begin{document}
\maketitle

\begin{abstract}
In both research and industrial settings, it is often necessary to expand the input/output channels of measurement instruments using relay-based multiplexer boards. In research activities in particular, the need for a highly flexible and easily configurable solution frequently leads to the development of customized systems.
To address this challenge, we developed a system optimized for automated direct current (DC) measurements. The result is based on a  $4\times4$ switching platform that simplifies measurement procedures that require instrument routing. The platform is based on a custom-designed circuit board controlled by a microcontroller. We selected bistable relays to guarantee contact stability after switching. We finally developed a system architecture that allows for straightforward expansion and scalability by connecting multiple platforms. We share both the hardware design source files and the firmware source code on GitHub with the open-source community.
This work presents the design and development of the proposed system, followed by the performance evaluation. Finally, we present a test of our designed system applied to a specific case study: the DC analysis of complex resistive networks through multi-point resistance measurements using only a single voltmeter and current source.
\end{abstract}

\section{Introduction}
Automated electrical measurement systems play a critical role in both scientific research and industrial environments, where rapid, repeatable, and scalable acquisition of electrical parameters is often required. In particular, the ability to route signals dynamically among multiple contact points, without manual rewiring, is essential in applications such as sensor array characterization \cite{massey2021amps,li2016real}, material characteristics studies \cite{pandey2018fully,porzuczek2021eit}, reliability testing of electronic components, and large-scale laboratory automation \cite{clark2021multiplexed}. In practice, this type of measurement requires a relay-based switching architecture (or switching matrices) \cite{jassemnejad2015prism}, allowing a flexible reconfiguration of measurement pathways while preserving electrical isolation, low contact resistance, and long-term stability. Even in presence of commercially available switching platforms \cite{keithley2006switching}, the research environments often require customized, open, and highly flexible systems that can be integrated into bespoke setups, support low-noise measurements, or be expanded to accommodate complex multi-node experiments \cite{lin2025highthroughput}.

In this work, we developed and tested a scalable $4\times4$-contacts relay switching platform optimized for automated DC resistance measurements. The system is built around a custom-designed printed circuit board (PCB) equipped with bistable relays, selected for their negligible static power consumption, high contact stability, and mechanical robustness \cite{standex2025reed}. In addition, the board can be used  also with solid-state relays, simply by mounting these in place of the electromechanical relays, given the fact that the footprints of both electromechanical and solid-state relays overlap. This also makes the board relevant for power applications; this version of the board operates with alternating current (AC) inputs.
A microcontroller manages the relay control logic and the communication interface, enabling seamless integration with external instruments and laboratory software. The system architecture is inherently modular: multiple units can be cascaded to create larger switching matrices, thereby supporting experiments which require elevated number of contact points for measurements \cite{abid2025multiplexed}.

The platform is intended as an open hardware resource: all design files—schematics and firmware—are available \cite{RMboard_git} freely online, allowing straightforward replication and modification. From a practical standpoint, the system reduces the overhead associated with reorganizing cables or reconfiguring setups between measurement steps. This characteristic makes it well-suited for tasks such as high-throughput characterization \cite{lin2025highthroughput}, multi-terminal probing of novel materials \cite{barnard2022feedback}, and studies requiring automated mapping of resistive structures.

This work provides a comprehensive description of the design, implementation, and performance of the proposed switching platform. In Section~\ref{sec:hardware} the board design is described and both schematic, layout and communication protocols are discussed. Then, in Section~\ref{sec:results}, the board performances are analyzed, first by characterizing the relay switching dynamics, including average latency and jitter, 
second, we evaluate the noise contribution introduced by the switching board when used in series with a nanovoltmeter. 
Finally, in Section~\ref{sec:ExperimentalValidation}, we propose an experimental validation of the use of the designed board by characterizing an 8-nodes resistive network connecting two relay matrices in cascade. 

\section{Relay Board Design and Architecture}\label{sec:hardware}
This project aims to provide a flexible switching platform that allows any input to be routed to any output. We opted for a single-board solution for  $4\times4$ board rather than a modular architecture to obtain a compact, ready-to-use instrument and to minimize additional connectors that could introduce noise. The resulting PCB has a moderate footprint (\SI{22}{cm} × \SI{10}{cm}) while offering full reconfigurability.

\begin{figure}[H]
    \centering
    \includegraphics[width=0.5\linewidth]{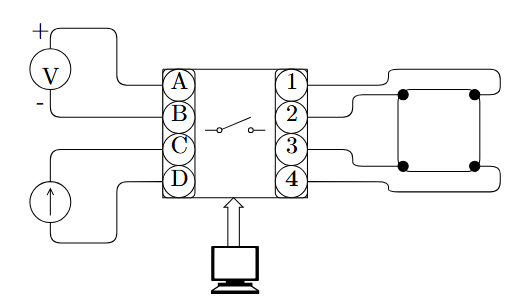}
    \caption{Example setup for the van der Pauw method. The voltmeter positive and negative terminals are connected, respectively, to A and B inputs, while the current generator source and sink terminals are, respectively, connected to C and D inputs. Closed connections can be described by a list of pairs composed of a letter(input) and a number(output), i.e. A3 B2 C1 D4. }
    \label{fig:hw_vdp_setup}
\end{figure}

As an example of use, the board can support standard four-probe configurations, including those required by the van der Pauw technique \cite{philips1958method} for sheet-property characterization (see \figurename~\ref{fig:hw_vdp_setup}). In this context, the relay matrix allows rapid switching between the measurement geometries needed to extract quantities such as sheet resistance or Hall voltage, without manually rewiring the sample. Although the van der Pauw method is a convenient illustration, the same routing capabilities enable many other layouts, such as eight-probe anisotropy measurements \cite{esposito2000determination}, alternative Hall geometries, or general purpose test setups. Thanks to the abstract input–output notation of the matrix, each configuration can be described unambiguously and recalled programmatically, allowing automated measurement sequences and complex protocols to be executed with minimal overhead.

In the following, we shall give an overview of the design of the circuit board, its working principles, and its connection with the NUCLEO board \cite{nucleoBoard}.

\subsection{Design}

\begin{figure}[H]
    \centering
    \includegraphics[width=0.9\linewidth]{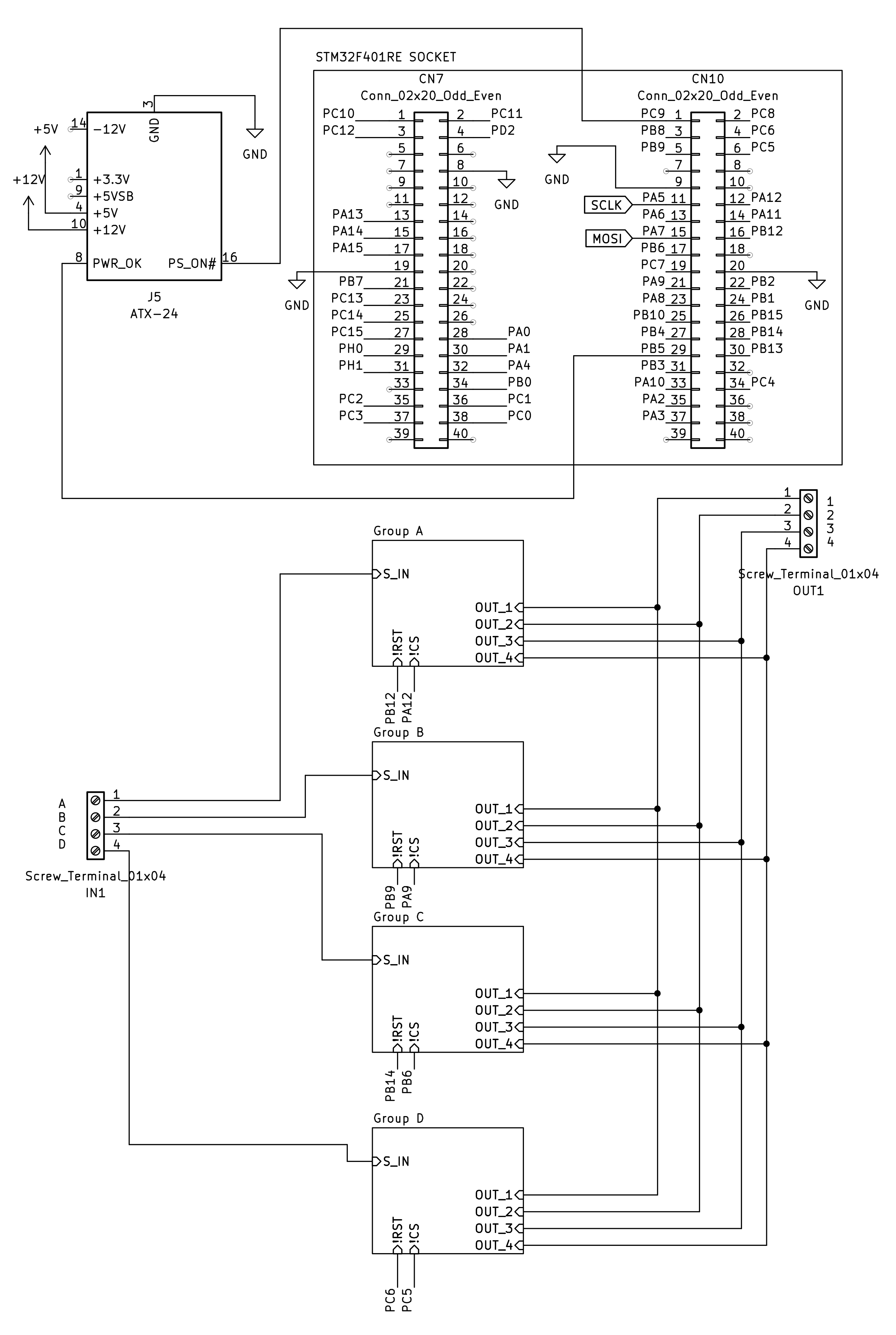}
    \caption{Main schematic of the relay board}
    \label{fig:main_schematic}
\end{figure}

The board adopts a hierarchical structure where each relay group is a 'subsheet' of the main schematic. This allows to represent the schematic of the board synthetically, shown in \figurename~\ref{fig:main_schematic}.   On the upper part there are the headers that provide the connection to the NUCLEO board, the ATX power supply, and the various relay groups organized as subsheets. Each subsheet is connected to one of the relay board input terminal blocks labeled A, B, C, D and to all the outputs labeled 1, 2, 3, 4.
The labels on the \SI{2.54}{mm} female headers follow the NUCLEO board pin numbering, and the ATX connector supplies the \SI{12}{V} for exciting the coils and powering the relay drivers. The \texttt{PS\_ON} pin is intended to turn on/off the ATX power supply, while the \texttt{PWR\_OK} reports that all voltages of the power supply have stabilized. The selected driver is the TPL9201 integrated circuit. It is an eight-channel low-side driver designed for controlling inductive or resistive loads such as relays, solenoids, or actuators. It integrates current-limited output stages, protection features, and an internal \SI{5}{V} regulator, enabling direct interfacing with microcontroller logic. Its SPI control interface minimizes pin usage while providing reliable switching of multiple channels.  

\begin{figure}[H]
    \centering
    \includegraphics[width=1\linewidth]{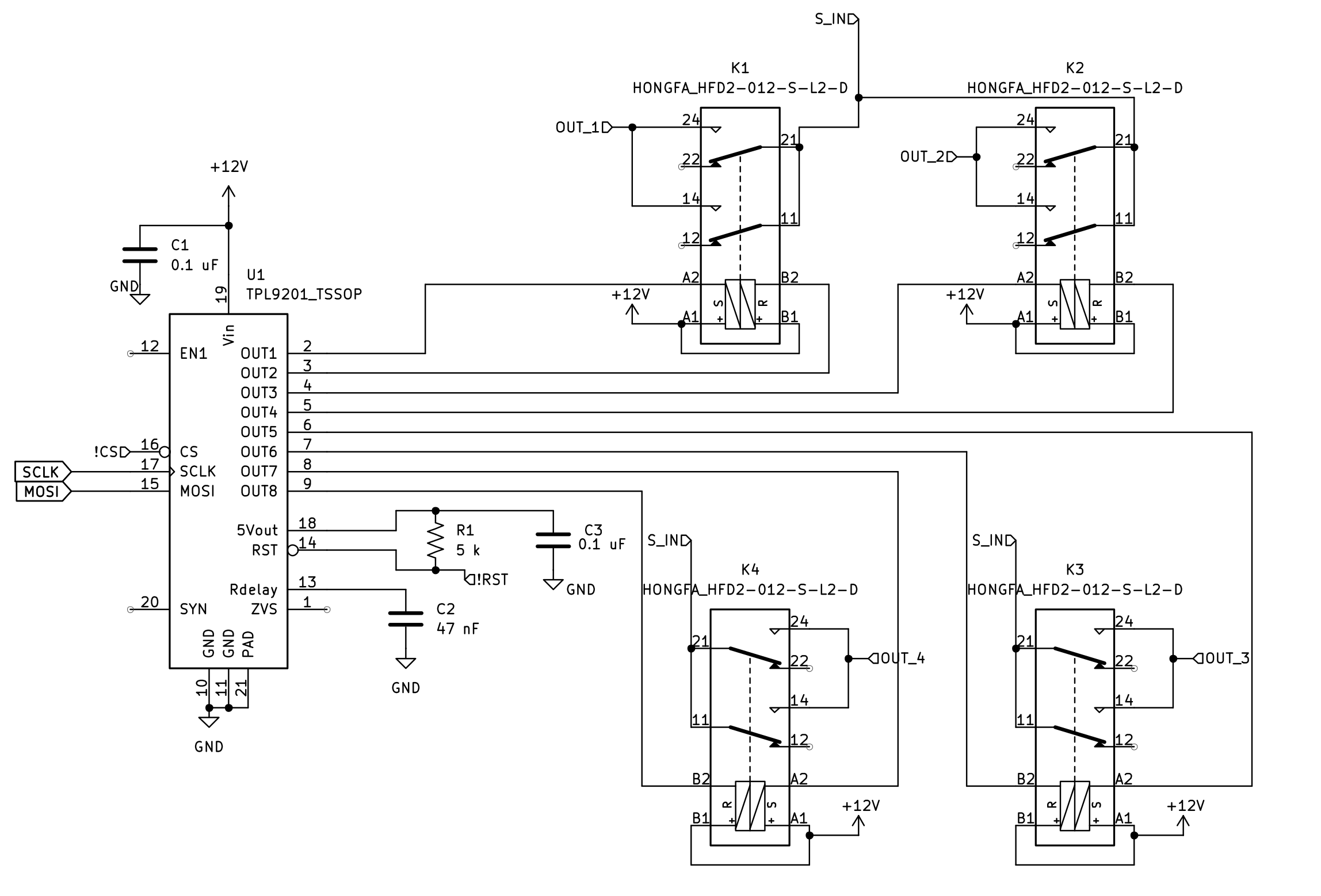}
    \caption{Schematic of the sub sheet. The board has been laid out on four instances of the latter. All commercially available models of the HFD2-012-x-L2-x relays with dual coil and \SI{12}{V} voltage are compatible with the board. }
    \label{fig:RelGroup}
\end{figure}

Each relay group (see \figurename~\ref{fig:RelGroup}), particularly each TPL9201, has a \texttt{!RST} pin; this pin can be either used for resetting the driver (in particular its stored status byte) by pulling down the associated GPIO, but also reports if the TPL9201 is ready to accept SPI data if high. This pin gets sensed before each commute and, if it is low, an error gets reported to the operator; this generally happens when the board is not properly powered. Each subsheet (\figurename~\ref{fig:RelGroup}) illustrates the connections between the TPL9201 driver and the four Hongfa HFD2/012-S-L2-D relays. Since the HFD2 is a bistable dual-coil relay (with separate set and reset coils), each TPL9201 driver controls four relays, corresponding to a total of eight coils.

The PCB has 4 layers and it is built upon FR4 substrate. The stack-up is as follows:
\begin{enumerate}
    \item SPI traces, GPIO and 12V supply
    \item SPI Traces and GPIO
    \item Commuted signals
    \item Commuted signals
\end{enumerate}

\begin{figure}[H]
    \centering
    \includegraphics[width=1\textwidth]{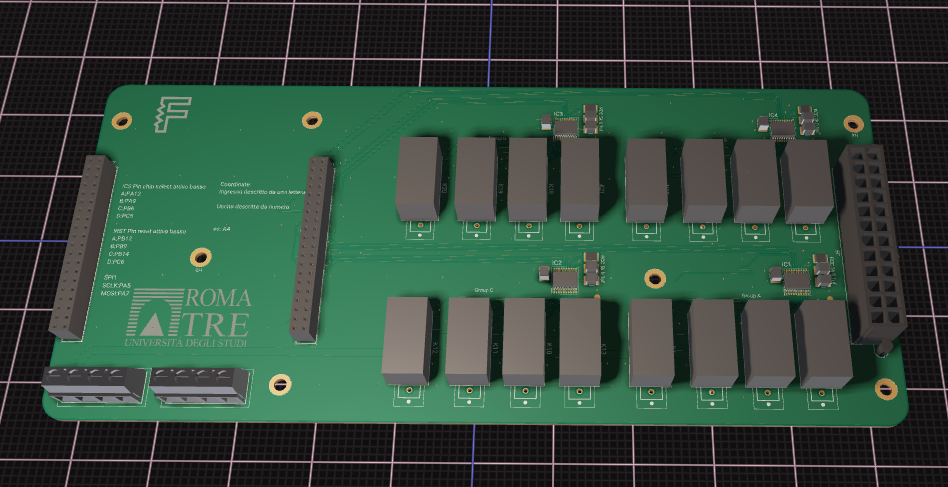}
    \caption{Front view of the assembled Relay Matrix board}
    \label{fig:BoardFrontView} 
\end{figure}

Each layer has a ground pour with several ground stitching vias across the board.
On top of the board, starting from the left (see Figure \ref{fig:BoardFrontView}), there are two different pairs
of connectors. The vertically oriented ones are 2x20 pin female connectors with a pitch of \SI{2.54}{mm}. 
The horizontal ones are two 4-pin screw terminals with a pitch of \SI{5.08}{mm}. Both kinds of connectors have through-hole terminals.
The NUCLEO board shall be inserted into the \SI{2.54}{mm} headers. In addition the matrix board also has a few mounting holes.
Then the relays are split into four areas. Each area contains four relays, a TPL9201 driver circuit, and various
capacitors/resistors that serve the purpose of bypass/pull-up. The footprints of the electro-mechanical (EMR configuration) and solid-state relays (SSR configuration) overlap, 
allowing one to choose which kind shall be assembled. On the far right of the board, there is a 24-pin ATX power supply connector, 
which also allows the monitoring of \texttt{PWR\_OK} (voltages stabilized) and \texttt{PS\_ON} (on/off ATX power supply) pins, whose purpose was explained earlier, that are specific to the ATX standard.
Ideally, a single ATX power supply shall be used for several matrices. 

The second layer serves routing purposes; it allows a few SPI-related pins to reach the right side connector (same orientation in figure {\ref{fig:BoardFrontView}}). In fact, the left side connector only provides additional ground pads. 
The remaining layers have the purpose of connecting the inputs from the left side screw terminal (A-D) to all the 16 relays, and the outputs of the relays to the 
right side terminal (1-4). The width of these tracks is \SI{2}{mm} and allows for currents up to \SI{2}{A}. No additional
vias have been added, thus layer transitions can only take place at through-hole pads.

\begin{figure}[H]
    \centering
    \includegraphics[width=1\textwidth]{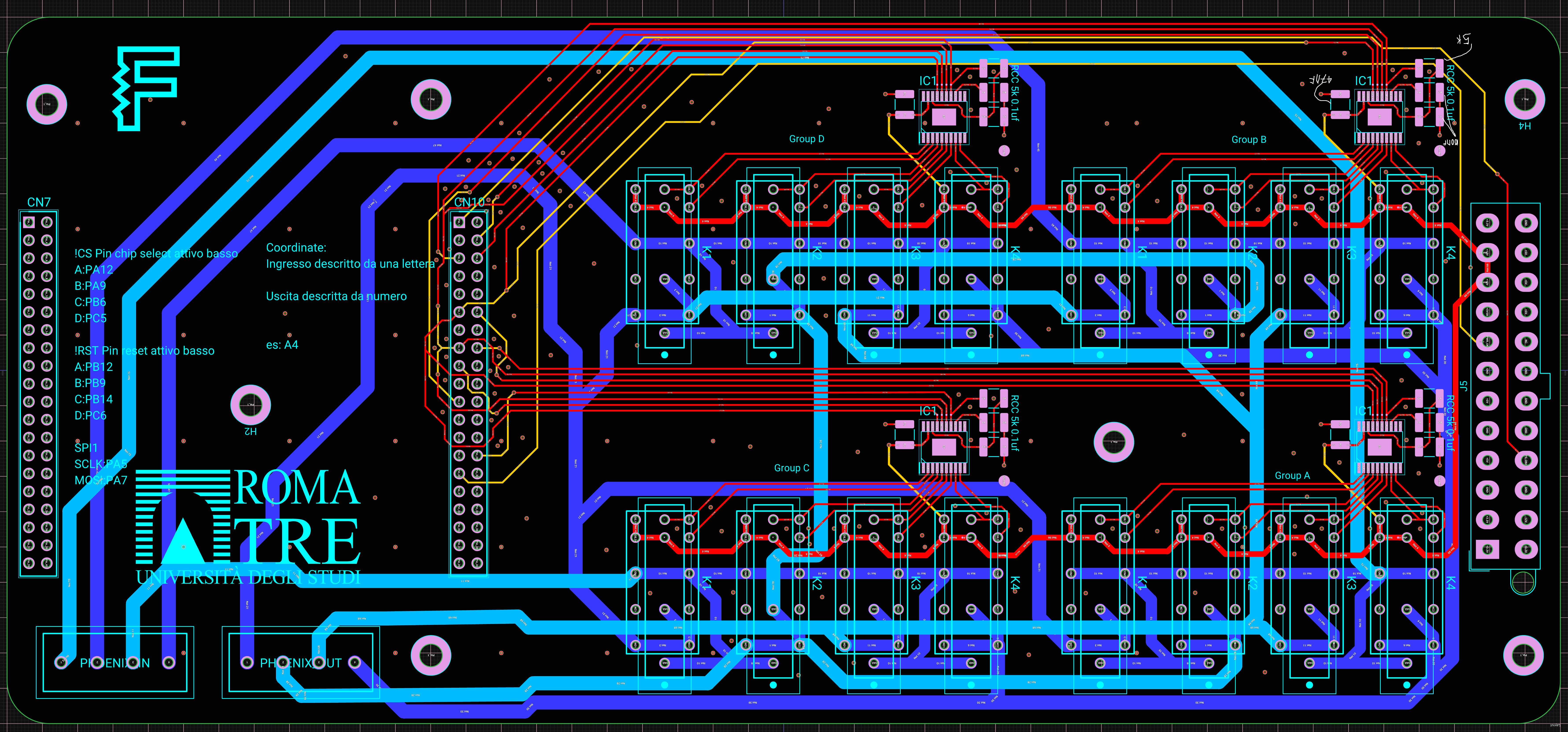}
    \caption{Complete view of the 4 layers}
    \label{fig:AllLayerView} 
\end{figure}

Due to the low-frequency-oriented design of the presented device, 
the lack of  impedance control, nor length matching on the signal traces, makes the board suitable only for signal frequencies from DC to \SI{60}{Hz}.

\subsection{Relay driving}
The operation of the relay control system relies on precise communication between the microcontroller and the TPL9201 driver. The driver manages an array of NMOS transistors which, in turn, control the relays. Understanding how data bytes are transmitted, interpreted, and translated into relay actions is essential for ensuring correct and safe operation, especially considering the differences between solid-state relays (SSR) and electromagnetic relays (EMR). The following section details the byte-level communication, the polarization requirements for bistable relays, and the implications for power management and persistence of relay states.

On each commute, a new byte of data gets sent to the TPL9201. The TPL9201 has a 1 byte buffer which allows to control an array of 8  NMOS transistors in an open-drain configuration. In the SSR version, only half of the byte actually carries
the relay group 'status', which means that 4 transistors of the TPL9201 array are left unused. In the latter, the 'set' transistors will stay on until a reset is issued, leading to additional power dissipation.
This mode of operation is said to be 'monostable'.
On the contrary, in the EMR configuration, the whole array is used, implying that the whole byte is relevant.  Each pair of coils (so-called set and reset coils) belongs to a relay. These pairs must be polarized in an exclusive way
in order to avoid damaging the component. When a group status change gets issued, each relay can be set, reset or left unchanged, this happens by updating the relevant bits of the TPL buffer; in this case an even index bit set to one means a set, while an odd index one implies a reset (counting from the LSB). 
After a certain amount of time specified from the HFD2 datasheet, which grants that the commutation is completed, 
the whole transistors array is switched off. While in the SSR configuration this would have resulted in an entire group being reset, in EMR configuration the relays are persistent in both status. This means that the board \SI{12}{V} supply may be switched off using \texttt{PS\_ON} pin and all the connections would persist. In the following, we will focus on the EMR configuration.

\begin{figure}[H]
    \centering
    \includegraphics[width=1\textwidth]{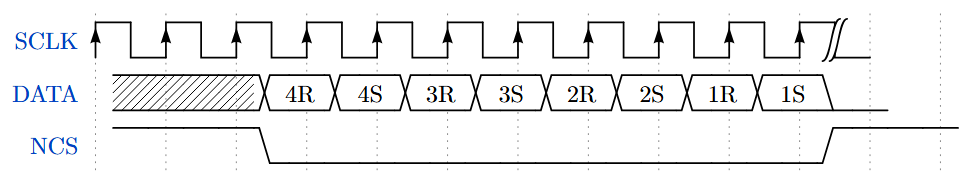}
    \caption{An overview of the SPI communication between the microcontroller and a single TPL9201. For each relay $\mathrm{R+S}\leq 1$.}      
    \label{fig:StatusByte} 
\end{figure}

In figure \ref{fig:StatusByte} there is an example update of a specific group. While the NCS signal selects a specific TPL, DATA will determine which relays have to be set (S) or reset (R). The first number identifies a specific relay (1-4), the second letter specifies which coil will be energized (set or reset). Each pair of bits of the said byte must have at least one zero in order to avoid polarizing simultaneously the two coils,
damaging the relay. 

After the latching time, NCS is pulled down and a null byte gets sent in order to turn off the whole transistors array; as the relays are bistable, there is no need to keep the coils energized. 

\subsection{Firmware}
The firmware of the STM32F401RE board, which controls the matrix, is based on the Hardware Abstraction Layer (HAL). Its aim is both to send data to all TPL9201 and provide
a basic serial interface which is heavily inspired by SCPI (Standard Commands for Programmable Interface) syntax.
The microcontroller exposes a UART port using the only USB connector the board has as a COM port, making any modern operating system capable of detecting it.
Each received command is stored in a memory area using a Direct Memory Access (DMA) peripheral, then an interrupt event is generated and subsequently served by the microcontroller.
The interrupt callback ends when the command string gets added to a linked list.
The finite state machine (FSM) loop checks during each iteration if there are any commands that are waiting to get processed. 
If the list is not empty, then the command in its raw form gets parsed by the parser component and a command structure gets associated to it, this instance gets put 
in another list. The interpreter will then check if the command is valid and eventually execute the associated action. 

\begin{figure}[H]
    \centering
    \includegraphics[width=1\textwidth]{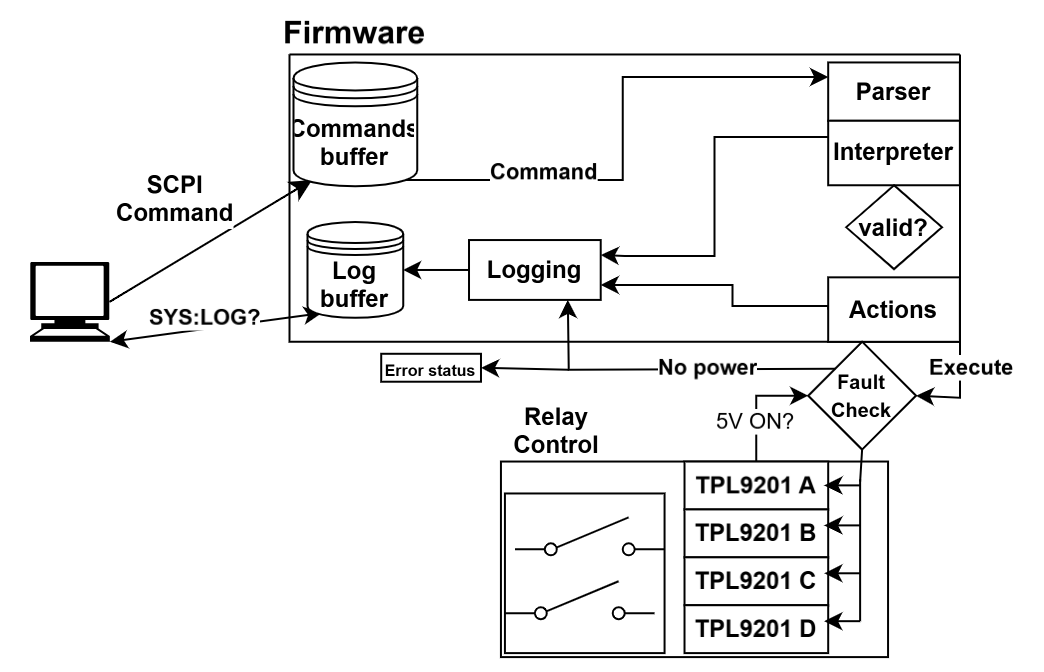}
    \caption{Flow chart of the firmware.}
    \label{fig:FlowChart} 
\end{figure}

After the execution, a log string gets populated with basic info on whether the action was successful, data returned, errors, etc. 
The contents of the latter may be obtained with a dedicated query. 
In figure \ref{fig:FlowChart}  the flow chart of the firmware operation is shown.

\section{Performance characterization}\label{sec:results}

In this section, we present the experimental characterization of the relay switching board that we have developed. The aim is to provide a comprehensive description of its key features and to evaluate its suitability for accurate measurements. In particular, we focused on three main aspects: (i) the switching dynamics of the relays, including their average latency and temporal jitter, (ii) the noise contribution introduced by the board when used in conjunction with a nanovoltmeter, and (iii) the demonstration of the modular nature of the system by connecting two boards in cascade and performing measurements on a complex resistive network. These results illustrate not only the performance of the device in terms of speed and noise, but also its versatility for different experimental configurations.

\subsection{Relay Switching Times}
\begin{figure}[H]
    \centering
    \includegraphics[width=1\textwidth]{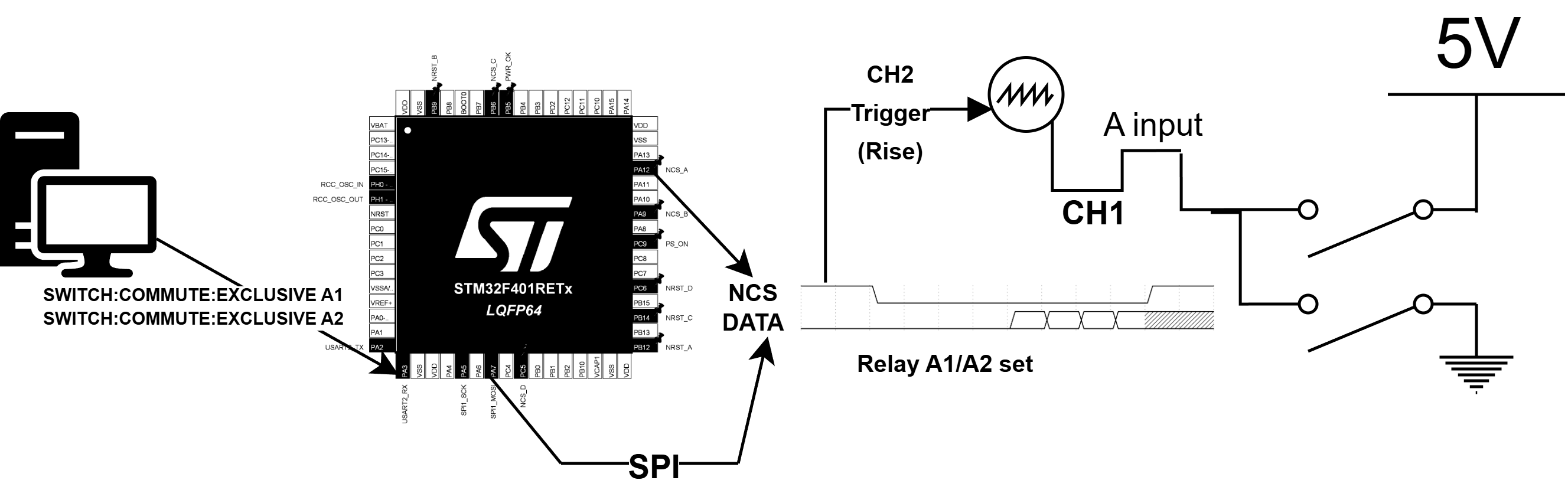}
    \caption{Setup used for measuring switching times. The NCS will select the TPL9201 of the group A, and subsequently send via SPI the byte needed for setting the A1/A2 relay. Oscilloscope waveform measurement will be triggered on the rising edge of the NCS signal. On each commute, the other relay will also be reset contemporarily.}
    \label{fig:transient_meas}
\end{figure}
The switching performance was investigated by applying control pulses from the microcontroller to the relay matrix, while monitoring the response with a digital oscilloscope. The test setup consisted of a 5~V DC source connected to the matrix inputs A and B, and the relay under test was used to connect the input channel of the oscilloscope to the 5~V level or to the reference one. Thus, we measured the switching times during both the 5~V to 0~V transition and the opposite one. In order to characterize the delay of the relay and not of the communication channel between the control pc and the board, the microcontroller output signal was connected to the second channel of the oscilloscope, and the trigger launched on this. In this way, we obtained a precise hardware trigger that defined the reference time of each commutation event.

A total of 1000 opening/closing cycles were acquired, yielding a dataset sufficiently large to construct both the average transient responses and the statistical distributions of the switching delays. The rise time was calculated when the curve crossed $1\;\mathrm{V}$, while the fall time was calculated at $4\;\mathrm{V}$. Both values were extrapolated using a piecewise linear fit on the waveform.

\begin{figure}%
    \centering
    \subfloat[\centering]{{\includegraphics[width=0.45\textwidth]{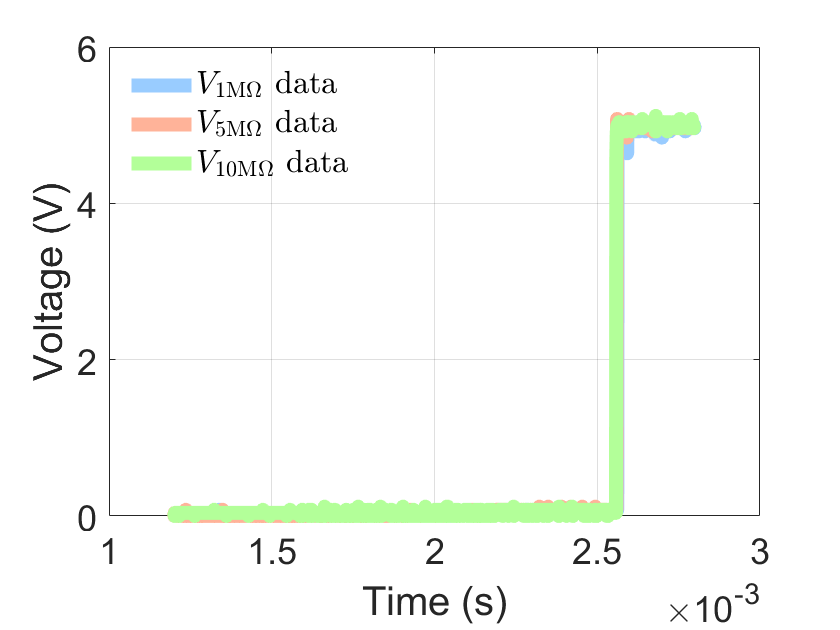} }}%
    \qquad
    \subfloat[\centering]{{\includegraphics[width=0.45\textwidth]{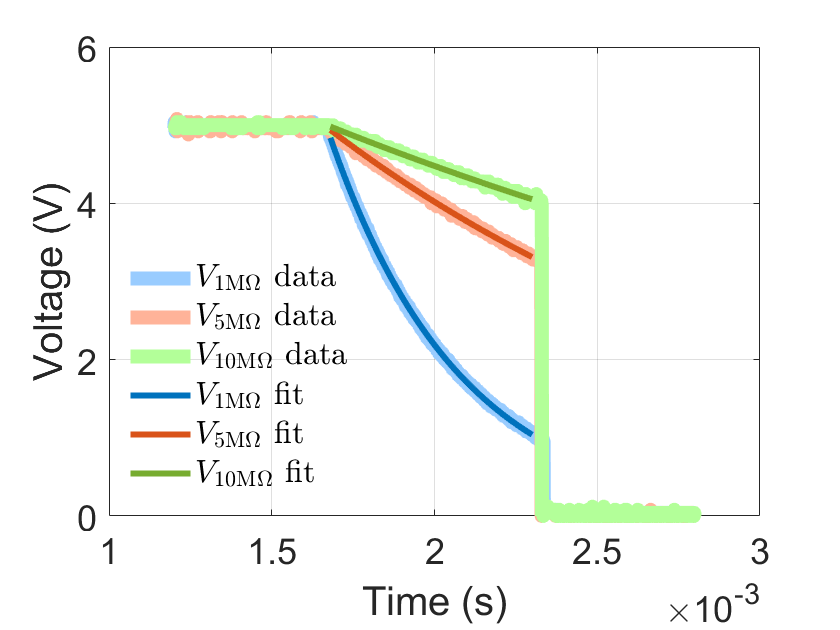} }}%
    \caption{(a) Transient during relay transition $0\;\mathrm{V}\rightarrow5\;\mathrm{V}$. (b) Transient during relay transition $5\;\mathrm{V}\rightarrow0\;\mathrm{V}$.}%
    \label{fig:transients}%
\end{figure}

The rising transition $0\;\mathrm{V}\rightarrow5\;\mathrm{V}$  showed the expected step-like behavior, characterized by a relatively sharp transition to the high state (see \figurename~\ref{fig:transients}~(a)). In contrast, the falling transition $5\;\mathrm{V}\rightarrow0\;\mathrm{V}$ revealed a two-stage process: a slow initial decrease followed by a rapid drop to zero volts (see \figurename~\ref{fig:transients}~(b)). This effect could be attributed to parasitic capacitances in the relay contacts during intermediate contact states. Similar two-stage opening transients have been discussed in the literature: the initial slow voltage decay can be explained by the discharge of stray capacitances associated with the open contact and wiring, which produce an effective RC time constant, while subsequent fast collapse corresponds to the final physical separation of the contacts \cite{Wu2019_Energies,Jang2022_Energies}. Previous analyses and application notes report comparable behavior and recommend accounting for distributed capacitances when interpreting switching transients \cite{SEL_ParasiticCapacitance}. Hence, the first part of the decay shown in \figurename~\ref{fig:transients}~(b) was interpreted with a RC decay model. To verify this interpretation, measurements were performed by changing the input resistance $R_{in}={\{1,5,10}\}\;\mathrm{M}\Omega$ of the oscilloscope used for the data acquisition. It is possible to see in \figurename~\ref{fig:transients}~(b) that as $R_{in}$ is increased the characteristic time $\tau=RC$ of the first stage of the discharge curve also increases. Hence, the first part of the discharge curves was fitted with a standard exponential model $V(t)=V_0e^{-\frac{t}{\tau}}$ to determine $\tau$. From this,  we obtained a parasitic capacitance $C\approx350$~pF. This is a plausible value for a distributed capacitance in the measurement system used. Whereas, for the $0\;\mathrm{V}\rightarrow5\;\mathrm{V}$ transitions shown in \figurename~\ref{fig:transients}~(a), we did not observe any significant variation resulting from the change in the oscilloscope input resistance $R_\mathrm{in}$. This does not imply that intermediate states of the relay did not occur during these transitions; rather, they could be interpreted as a parasitic capacitances that are already discharged by this time, so no measurable effect can be detected.

\begin{figure}%
    \centering
    \subfloat[\centering]{{\includegraphics[width=0.45\textwidth]{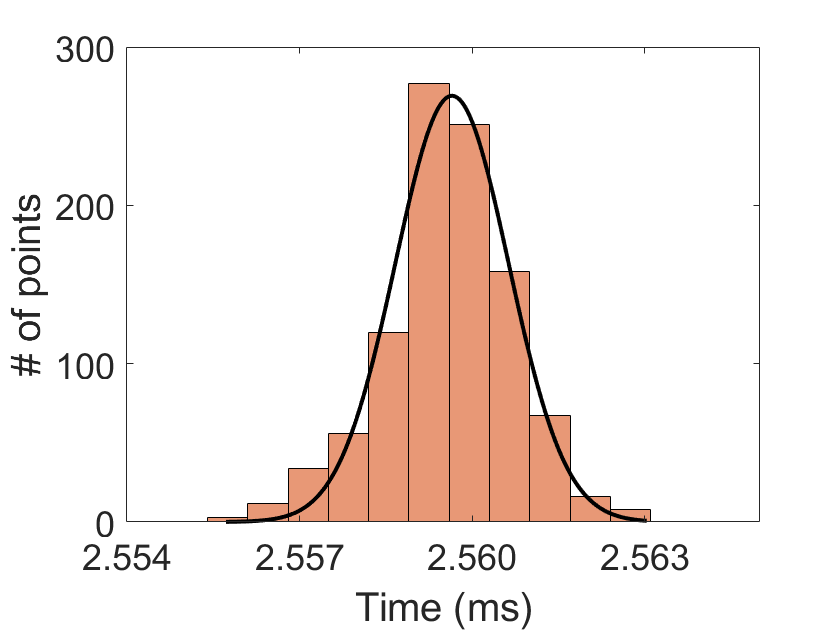} }}%
    \qquad
    \subfloat[\centering]{{\includegraphics[width=0.45\textwidth]{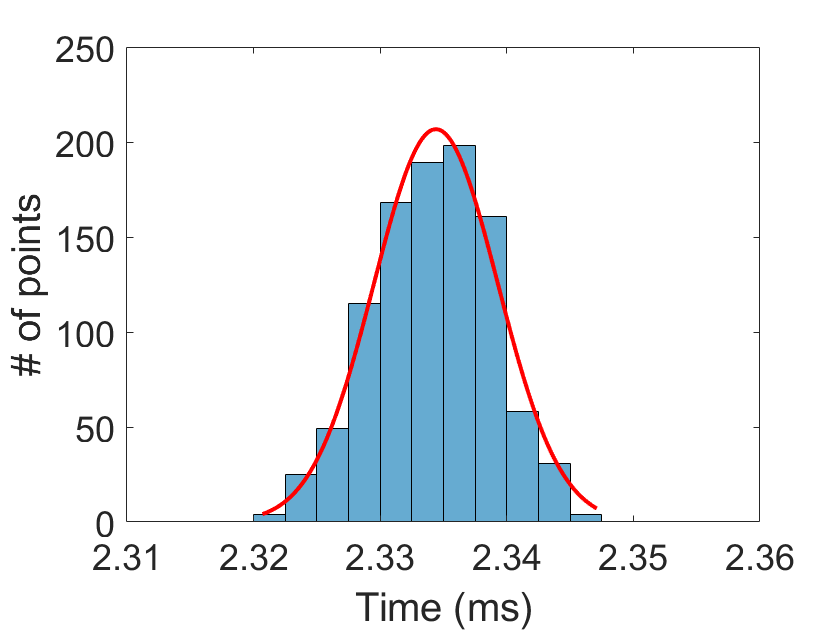} }}%
    \caption{Histograms of the switching times during the transition $0\;\mathrm{V}\rightarrow5\;\mathrm{V}$ in (a), and the transition $5\;\mathrm{V}\rightarrow0\;\mathrm{V}$ in (b). }%
    \label{fig:transientsAnalysis}%
\end{figure}

Performed statistical analysis of the commutation times revealed different behaviors in commutation types (\figurename~\ref{fig:transientsAnalysis}).  For the rising commutation, the mean time is set at $t_{r}\approx2.559$~ms and its standard deviation is $\sigma_{r}\approx1.136\times10^{-6}$~s. While for the falling commutation time, the mean value is $t_{f}\approx2334$~ms and its standard deviation is $\sigma_{r}\approx4.685\times10^{-6}$~s. 

\begin{figure}
    \centering
    \includegraphics[width=1\linewidth]{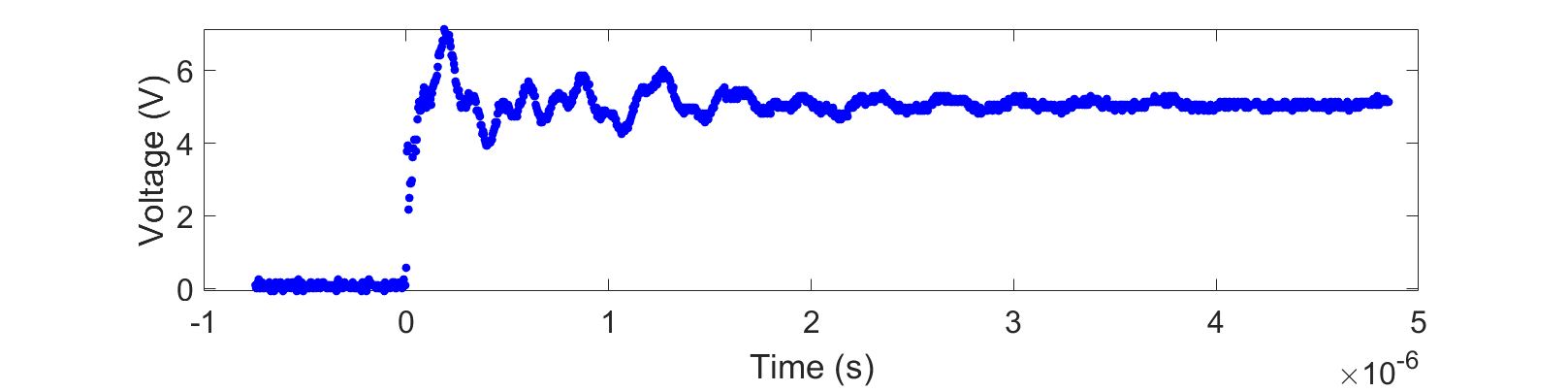}
    \caption{Voltages across the switch contacts during the commutation}
    \label{fig:OscCommut}
\end{figure}

Regarding the time required for the relay to stabilize after switching, individual transitions were sampled at a higher temporal resolution to capture the typical bouncing behavior expected from a switch of this type. An example is shown in \figurename~\ref{fig:OscCommut}, where oscillations caused by contact bounce following the transition are clearly visible. From this, a settling time of approximately $t_s\simeq4\,\mu\text{s}$ can be estimated, which is notably short when compared to both the switching time itself and the associated jitter. Therefore, this settling time can be considered negligible for practical applications.

\subsection{Noise Performance}

The noise introduced by the relay board was assessed using a nanovoltmeter (NV), Agilent 34420, configured at its most sensitive scale (10~mV full range). A stable DC source, consisting of a battery followed by a voltage divider, was used to provide the input signal. Two sets of measurements were acquired: (i) directly connecting the NV to the source, and (ii) inserting the relay board in series between the source and the NV. The integration time was set to 1~NPLC in order to optimize the trade-off between noise rejection and acquisition rate.

To ensure precise timing control, we instructed the NV to acquire 1000 samples internally and to transfer the entire dataset in a single operation, rather than streaming individual values. This method guaranteed stable sampling intervals, which were verified by analyzing batches of increasing size (100, 200, 300 samples, etc.) and fitting the corresponding time intervals with a linear function. The slope of this fit function provided an accurate estimate of the sampling period $t_s\approx46.1$~ms.

\begin{figure}
     \centering
     \includegraphics[width=0.6\linewidth]{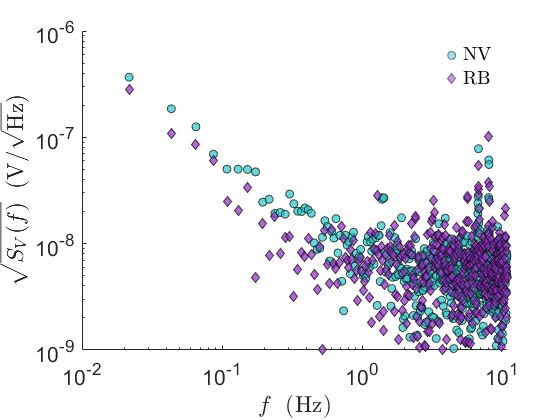}
     \caption{Noise amplitude spectral density $\sqrt{S_V}$ measured with only the nanovoltmeter (cyan circles), and with the relay board (purple rhombus). }
     \label{fig:NoiseComparison}
 \end{figure}

 To analyze the noise, we computed the amplitude spectral density $\sqrt{S_V}$ via fast Fourier transform (FFT) in \figurename~\ref{fig:NoiseComparison}. The comparison of the spectra obtained with and without the relay board showed excellent agreement across the entire frequency range. In particular, the $1/f$ component dominated up to about 1~Hz, while the white-noise plateau was observed at higher frequencies. The white-noise level is compatible with the NV characteristics. No additional spectral features or excess noise were introduced by the relay board, demonstrating that the switching elements do not compromise the intrinsic performance of the NV, at least within the noise limits of our setup.

\section{Experimental validation}\label{sec:ExperimentalValidation}

\begin{figure}[H]
    \centering
    \includegraphics[width=0.35\linewidth]{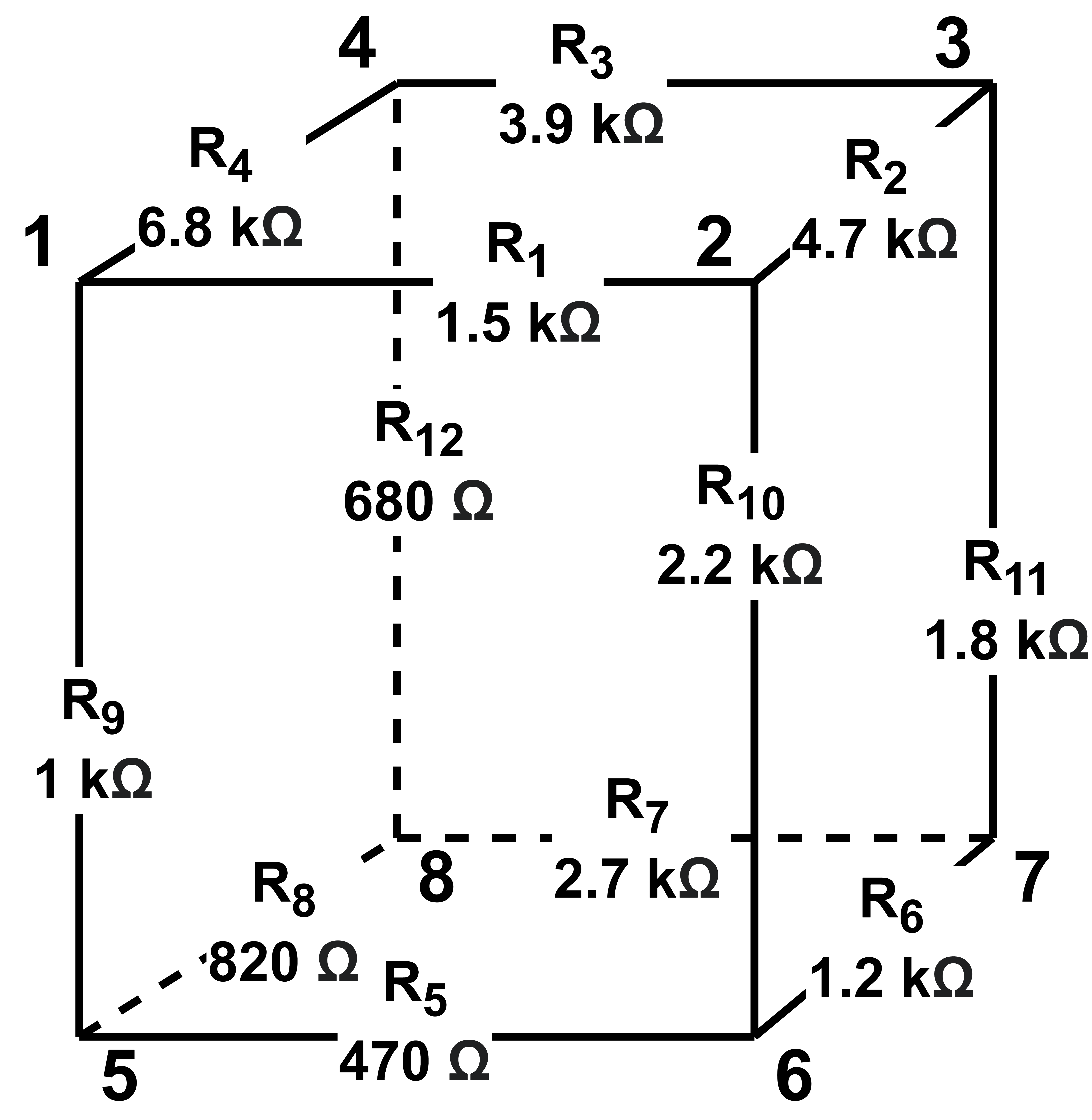}
    \caption{Three dimensional representation of the resistors cube. Each vertex identifies a node of the associated graph. Each edge has an associated value of resistance(thus conductance).}
    \label{fig:rcube}
\end{figure}

To demonstrate the modularity and scalability of the switching system, we cascaded two 4$\times$4 relay boards, effectively producing an 8-channel reconfigurable switching matrix. As a testbench, we constructed a three-dimensional resistive network in the form of a cube with twelve independently chosen resistances, one for each edge, as shown in \figurename~\ref{fig:rcube}. The routing system allows arbitrary selection of the current-injection pair and of the voltage-measurement pair, enabling a complete four-terminal characterization of the network without any manual rewiring.

A four-terminal measurement on an $N$-node resistive network is completely determined by: (i) a pair of nodes used for current injection and extraction, and  
(ii) a second pair of nodes used for voltage sensing.  In our case $N=8$.
Each pair is an unordered 2-element subset of the node set.  For the current terminals, the unordered nature follows from the fact that exchanging the two nodes merely inverts the sign of the injected current vector. When the voltage terminals are exchanged simultaneously, the measured differential voltage changes sign as well, leaving the corresponding four-terminal resistance
$ R_{AB;CD} = \frac{V_C - V_D}{I_0} $
unchanged in magnitude.  Thus, the measurement is physically identical under the exchange $A\leftrightarrow B$ and $C\leftrightarrow D$, and each terminal pair is combinatorially a set rather than an ordered tuple.
The number of admissible current pairs is the number of 2-element subsets of 8 nodes $N_{AB} = \binom{8}{2} = 28$. Once the current terminals $\{A,B\}$ are chosen, they cannot be reused as voltage terminals; hence the voltage pair must be chosen among the remaining 6 nodes $N_{CD} = \binom{6}{2} = 15$.
Since the two choices are independent, the total number of physically distinct four-terminal measurements is
\begin{equation}
N = N_{AB}\,N_{CD}
  = \binom{8}{2}\binom{6}{2}
  = 28 \times 15
  = 420.
\end{equation}

This enumeration counts only genuinely different measurements, excluding all permutations that lead to sign changes but not to new physical information.

The cube network consists of eight nodes and twelve resistive edges. The adopted numbering is shown in \figurename~\ref{fig:rcube}. 
Let $g_k=1/R_k$ denote the conductance of edge $k$. The nodal admittance matrix of the cube is an $8\times 8$ sparse symmetric matrix:

\begin{equation}
\scalebox{0.7}{$
\textbf{G} =
\begin{pmatrix}
g_1+g_4+g_9 & -g_1 & 0 & -g_4 & -g_9 & 0 & 0 & 0 \\
-g_1 & g_1+g_2+g_{10} & -g_2 & 0 & 0 & -g_{10} & 0 & 0 \\
0 & -g_2 & g_2+g_3+g_{11} & -g_3 & 0 & 0 & -g_{11} & 0 \\
-g_4 & 0 & -g_3 & g_3+g_4+g_{12} & 0 & 0 & 0 & -g_{12} \\
-g_9 & 0 & 0 & 0 & g_5+g_8+g_9 & -g_5 & 0 & -g_8 \\
0 & -g_{10} & 0 & 0 & -g_5 & g_5+g_6+g_{10} & -g_6 & 0 \\
0 & 0 & -g_{11} & 0 & 0 & -g_6 & g_6+g_7+g_{11} & -g_7 \\
0 & 0 & 0 & -g_{12} & -g_8 & 0 & -g_7 & g_7+g_8+g_{12}
\end{pmatrix}.
$}
\end{equation}

The nodal voltages follow from
\begin{equation}\label{eqn:nodalM}
\textbf{G}\,\boldsymbol{V}=\boldsymbol{I},
\end{equation}
where $\boldsymbol{I}$ is the current-injection vector. 
The twelve edge resistances of the cube are reconstructed from the complete set of 
420 four–terminal measurements.  
To guarantee positivity of the parameters and to improve the numerical conditioning of the inverse problem, the optimization is carried out using a logarithmic parametrization of the conductances:
$ x_k = \log(g_k), k=1,\dots,12. $
The logarithmic parametrization improves the robustness of the nonlinear fit by enforcing $g_k>0$ exactly and reducing the effective dynamic range of the parameters.
For a given configuration of current–injection nodes $\{A,B\}$ and voltage–sensing nodes $\{C,D\}$, the theoretical four–terminal equivalent resistance is computed from the nodal equation 
$\mathbf{G}(g)\,\mathbf{V}=\mathbf{I}$ (see Eq. \ref{eqn:nodalM}) as
\begin{equation}
f(x)^{(m)} =R_{AB;CD}=R_{eq}^{(m)}= \frac{V_C(g) - V_D(g)}{I_0}, \qquad m=1,\dots,420,
\end{equation}
where $I_0$ is the excitation current and $V_C(g)$, $V_D(g)$ are the voltages at nodes $C$ and $D$, obtained after solving the nodal system.

We used a Keithley 2400 SMU as a current source (set to $\pm 10\,$mA) and a Keithley 2000 DMM as a differential voltmeter for these validation measurements.  
Each measurement is acquired twice, once with $+I_0$ and once with $-I_0$, and the two values are averaged.  
This current–reversal technique suppresses thermoelectric offsets.

To reconstruct conductance values, we performed a minimization of a Tikhonov – regularized objective function \cite{willoughby1979solutions}:
\begin{equation}
\Phi(x)
= \big\| f(x) - y \big\|_2^2
+ \lambda_{\mathrm{reg}}^{\,2}\, \big\| x - x_0 \big\|_2^2,
\end{equation}
where $y \in \mathbb{R}^{420}$ is the vector of measured four–terminal resistances,
$x_0$ is a prior estimate and $\lambda_{\mathrm{reg}}$ is the regularization parameter that controls the weight of the prior estimate.  
After minimization, the fitted resistances are obtained as
\begin{equation}
\hat{R}_k = \frac{1}{\hat{g}_k} = e^{- \hat{x}_k}.
\end{equation}

\begin{figure}[t]
    \centering
    \begin{minipage}{0.48\linewidth}
        \centering
        \includegraphics[width=\linewidth]{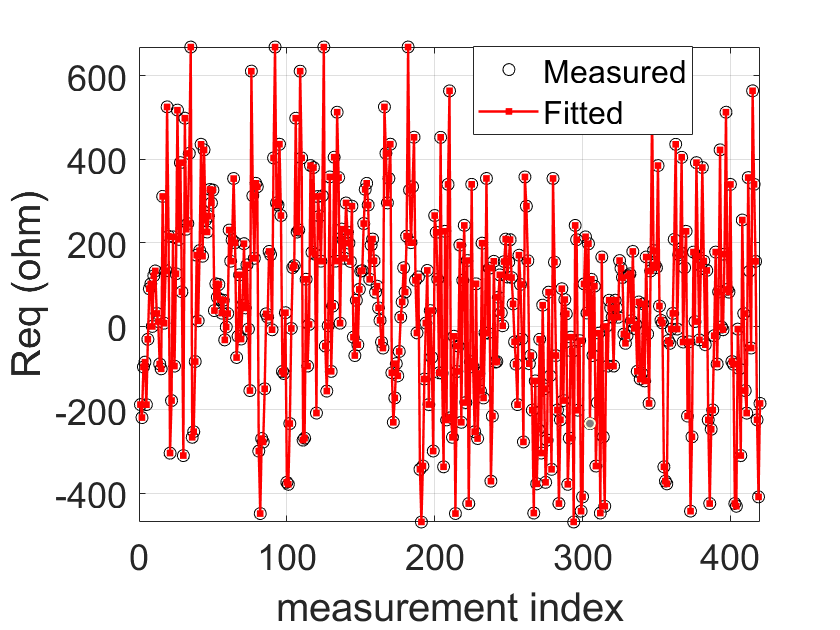}
        \\(a)
        \label{fig:FitRCuboa}
    \end{minipage}
    \hfill
    \begin{minipage}{0.48\linewidth}
        \centering
        \includegraphics[width=\linewidth]{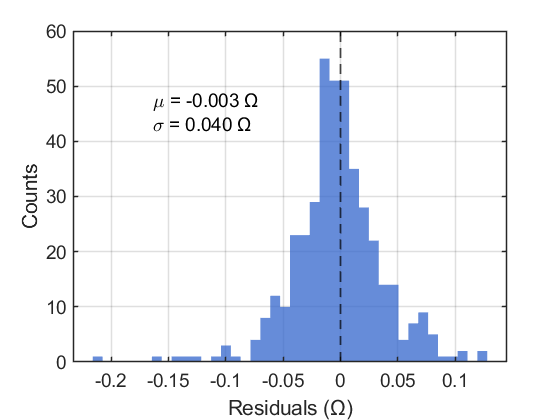}
        \\(b)
        \label{fig:ResidualsHistogram}
    \end{minipage}
    \caption{(a) Measured and fitted values of the 420 equivalent resistances. 
    (b) Histogram of the fitted $R_{\text{eq}}$ residuals.}\label{fig:FitRCubo}
\end{figure}

The comparison between the measured $y$ and fitted $f(x)$ values, together with the distribution of the residuals, are shown in \figurename~\ref{fig:FitRCubo}.

\section{Conclusions}

We demonstrate experimentally that the proposed relay-based switching platform is a viable solution for automated DC measurements requiring flexible routing, low noise, and modular scalability. The characterization of the switching dynamics shows a clear asymmetry between closing and opening transitions, but with timescales fully compatible with rapid automated acquisition. In particular, the closing time remained in the range of 2–4 ms, while opening transitions exhibited a slightly longer effective electrical settling of 5–7 ms, consistent with the presence of a small parasitic capacitance at the contacts. Fitting the transient response to a simple RC model yields an effective capacitance of approximately $350$~pF, in line with typical relay-contact and PCB parasitic contributions. Switching jitter remained below 200 µs, ensuring reliable timing even in fast measurement sequences.

The noise analysis revealed that the board introduces no measurable degradation of the nanovoltmeter performance within the experimental sensitivity. The preservation of both the low-frequency $1/f$ behavior and the high-frequency white-noise floor confirms that the chosen bistable relays and compact layout do not contribute additional pickup or dynamic noise beyond the intrinsic instrument limits.

The case study involving a three-dimensional resistive network demonstrated the practical scalability of the system. By cascading two units to form an 8-channel matrix, we performed a complete set of 420 four-terminal measurements without manual rewiring, enabling automated reconstruction of the network’s resistive distribution. This capability is essential in large-scale or high-throughput experiments where systematic multi-point acquisition is required.

Overall, the combination of low noise, predictable switching behavior, and inherent modularity makes the proposed system a powerful open-hardware tool for automated electrical measurements.

\bibliographystyle{unsrt}
\bibliography{biblio}

\end{document}